\documentstyle[12pt]{article}
\if@twoside  m
\else
\fi
 \newcommand{\ie}{{\em i.e.}}
 \newcommand{\eg}{{\em e.g.}}
 \newcommand{\etc}{{\em etc. }}
 \newcommand{\DD}{{\cal D}}
 \newcommand{\aleq}{\stackrel{<}{\scriptstyle\sim}}
 \newcommand{\ageq}{\stackrel{>}{\scriptstyle\sim}}

\newcommand{\cf}{{\em cf. }}

\newcommand{\R}{I\!\!R}

\begin{document}
\title{}
\date{}
\author{}
\noindent
{\Large\bf Evanescent modes in a multiple scattering \\ factorization}
\vspace{15mm}
\begin{quote}
{\large P.~Exner$^{a,b}$ and M.~Tater$^a$}
\vspace{3mm}

{\em a) Nuclear Physics Institute, Academy of Sciences \\
CZ--250 68 \v Re\v z near Prague \\ b) Doppler
Institute, Czech Technical University, \\ B\v rehov\'a 7,
CZ--115 19 Prague, Czech Republic}
\vspace{10mm}

We discuss differences between the exact S--matrix for scattering
on serial structures and a known factorized expression constructed
of single--element S--matrices. As an illustration, we use an
exactly solvable model of a quantum wire with two point impurities.
\end{quote}
\vspace{10mm}


\newpage

The core of the scattering--operator concept is that it relates
asymptotic states far away from the region where the interaction
affects the motion substantially. It happens often, however, that
the interaction is localized to a certain domain $\,\Omega\;$;
since solutions to the free equation are usually known, we may
rephrase then the scattering problem as a map between the solutions
at the boundary of $\,\Omega\,$ which can be continued (in some
direction) into the corresponding asymptotic states. If we want
to keep a terminological distinction, it is more appropriate
to speak about the {\em prescattering} operator in this case.

Such a ``finite--distance" scattering is particularly useful in
situations when the interaction support is a union of a finite number
of domains $\,\Omega_j\;$; it can help to solve the full problem by
means of scattering on the ``components". This subject has become actual
in connection with recent studies of quantum--wire superlattices
[1--13]
in which a number of elements, usually of the same type, is arranged
into a serial structure. The physically relevant quantity is the
conductivity which is related directly to the electron scattering
in the superlattice by Landauer's formula.

There are several ways how to derive the S--matrix components of the
serial structure, \ie, transmission and reflection amplitudes from
analogous quantities of a single element. The above sketched observation
applies directly if the number of linearly independent solutions
entering and leaving each component scatterer is finite. This is the
case, \eg, when the system in question is a graph \cite{SJ,TO}; it
may not be easy to find the solution interconnecting the scatterers
if an external field is applied, but the algebraic part of the problem
is well established \cite{ES,ESS}.

However, quantum mechanics lures always around ready to show who is
the master of our physical world. In general, diferent components of the
wave function, say, different transverse modes in a quantum wire, do
not cease to be correlated even out of the support of the interaction.
Here the difference between the scattering and prescattering operator
shows, because the sets of states the latter maps onto each other are
larger; in addition to true asymptotic states which live eternally
they contain also such that die out when the distance from the
interaction region increases.

It would be certainly worth to formulate properly relations between this
approach, which is inherently time--independent, and the rigorous scattering
theory \cite{AJS,RS}. Our aim here, however, is more practical and
connected with the mentioned studies of quantum--wire superlattices.
With few exceptions their authors include the evanescent states into
the iterative procedures of computing the S--matrix, so the result suffers
no theoretical defect. The weak point of all numerical calculations stems
from the necessity to restrict the used family of states to a finite number;
the stability aspect is usually handled by a vague observation that the
involved series converge fast enough. In this letter we want to discuss
this problem in more detail; we are going to derive explicit error
estimates in a solvable model in which the individual scatterers are point
impurities in a straight strip.

\section{S--matrix factorization}

Elements of a one--dimensional superlattice are linearly arranged and most
authors use various transfer--matrix modifications connecting solutions to
the left and right of a given scatterer
[1,2,5,9--11];
some admit numerical problems due to a fast growth of higher evanescent
modes. It has been argued recently \cite{Xu2} that the approach, which we
called pre--S--matrix, offers a more stable scheme relating instead the
ingoing and outgoing waves.

\begin{figure}
   \begin{picture}(120,80)
      \linethickness{2pt}
      \put(30,60){\line(1,0){350}}
      \put(30,20){\line(1,0){350}}
      \thinlines
      \put(205,5){\vector(0,1){70}}
      \put(10,20){\vector(1,0){390}}
      \put(35,20){\vector(0,1){40}}
      \put(35,60){\vector(0,-1){40}}
      \put(21,36){$d$}
      \put(210,71){$y$}
      \put(395,25){$x$}
      \put(136,30){\circle*{2.5}}
      \put(271,48){\circle*{2.5}}
   \end{picture}

\vspace{5mm}

\caption{A pair of scatteres in the strip.}
   \end{figure}
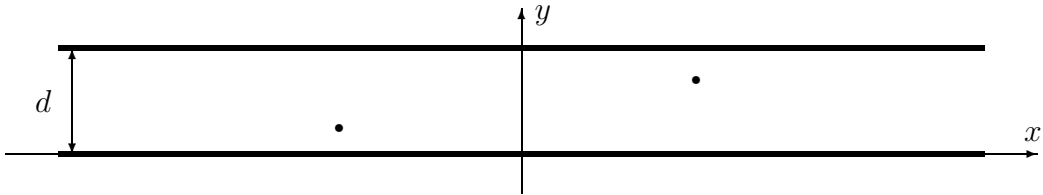

Consider therefore a pair of scatterers $\,S_1,\,S_2\,$ as sketched on
Figure~1, where $\,A_+\,$ is the family of solutions entering $\,S_1\,$
from the left {\em etc.;} in particular, if the corresponding external
motion is free, $\,A_+\,$ is a column vector $\,\left( a_{\ell}^{(+)}\,
{\rm e}^{{\rm i}k_{\ell}x}\right)\,$ with $\,k_{\ell}\,$ being the
channel momenta. The pre--S--matrices can be written in terms of the
reflection and transmission parts,
   \begin{equation} \label{S matrices}
\left( \begin{array}{c} B_- \\ B_+ \end{array} \right) \,=\,
\left( \begin{array}{cc} R_1 & \tilde T_1 \\ T_1 & \tilde R_1
 \end{array} \right)
\left( \begin{array}{c} A_+ \\ A_- \end{array} \right)\,, \quad
\left( \begin{array}{c} D_- \\ D_+ \end{array} \right) \,=\,
\left( \begin{array}{cc} R_2 & \tilde T_2 \\ T_2 & \tilde R_2
 \end{array} \right)
\left( \begin{array}{c} C_+ \\ C_- \end{array} \right)\,,
   \end{equation}
where the tilded quantities corresponding to the passage from the
right to the left are obtained from $\,R_j,\,T_j\,$ by mirror
transformation. The above form is convenient because the numbers of
involved modes at the two sides of $\,S_j\,$ may be different. Let us
stress that the pre--S--matrices in this setting need not be unitary.
The evanescent modes --- if included --- do not contribute to the
probability current, and the propagating ones enter with different
velocities: if the corresponding part of $\,S_j\,$ should be unitary,
we have to multiply the $\,(n,m)$--th amplitudes by $\,(k_m/k_n)^{1/2}$.

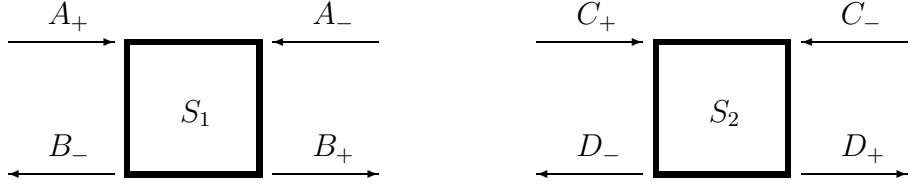
\begin{figure}
   \begin{picture}(120,80)
     \linethickness{2pt}
      \put(80,10){\line(1,0){51}}
      \put(130,10){\line(0,1){51}}
      \put(130,60){\line(-1,0){51}}
      \put(80,60){\line(0,-1){51}}
      \put(280,10){\line(1,0){51}}
      \put(330,10){\line(0,1){51}}
      \put(330,60){\line(-1,0){51}}
      \put(280,60){\line(0,-1){51}}
      \thinlines
      \put(35,60){\vector(1,0){40}}
      \put(75,10){\vector(-1,0){40}}
      \put(235,60){\vector(1,0){40}}
      \put(275,10){\vector(-1,0){40}}
      \put(335,10){\vector(1,0){40}}
      \put(375,60){\vector(-1,0){40}}
      \put(175,60){\vector(-1,0){40}}
      \put(135,10){\vector(1,0){40}}
      \put(100,30){$S_1$}
      \put(300,30){$S_2$}
      \put(50,67){$A_+$}
      \put(50,17){$B_-$}
      \put(150,67){$A_-$}
      \put(150,17){$B_+$}
      \put(250,67){$C_+$}
      \put(250,17){$D_-$}
      \put(350,67){$C_-$}
      \put(350,17){$D_+$}
   \end{picture}

\vspace{5mm}

\caption{A pair of point scatteres in the strip.}
\end{figure}

Our aim is to express $\,B_-,\,D_+\,$ from given $\,A_+,\,C_-\,$.
The relations (\ref{S matrices}) yield a set of four operator equation
which can be solved if we identify the left and right traveling
solutions between the scatterers, $\,D_-=A_-\,$ and $\,C_+=B_+\,$.
Using the identity
$$
I+X(I\!-\!YX)^{-1}Y\,=\,(I\!-\!XY)^{-1}
$$
which is valid whenever the inverses make sense, we find by a
straightforward algebra that the combined pre--S--matrix is
   \begin{equation} \label{factorized S matrix}
S_{12}\,=\,
\left( \begin{array}{cc}
R_1\!+\!\tilde T_1(I\!-\!R_2\tilde R_1)^{-1} R_2 T_1
& \tilde T_1 (I\!-\!R_2\tilde R_1)^{-1} \tilde T_2
\\ \\ T_2 (I\!-\!\tilde R_1 R_2)^{-1} T_1 &
\tilde R_2\!+\!T_2(I\!-\!\tilde R_1 R_2)^{-1} \tilde R_1 \tilde T_2
 \end{array} \right)\,.
   \end{equation}
The blocks of this operator acquire an illustrative meaning if
we expand the inverses into geometric series \cite{VOK1}. Sometimes
$\,S_2\,$ is obtained by shifting and/or mirroring another
scatterer, then one has to find the operator which represents
this transformation. This is equivalent to finding the general
solution in the intermediate region between $\,S_1\,$ and $\,S_2\,$
which may not be easy if the motion is not free there \cite{ESS}.

The operator (\ref{factorized S matrix}) describes the combined
scatterer exactly if the space into which $\,S_1\,$ and
$\,S_2^{-1}\,$ map is large enough to accomodate all generalized
eigenvectors of the corresponding Schr\"odinger equation. A truncation
induces an error which in general is not easy to estimate; we
shall do that below in an explicitly solvable model.

\section{Point--interaction scattering in a strip}

The model describes a nonrelativistic particle confined to straight
strip of a width $\,d\,$ with the hard--wall boundary subject to
point perturbations simulating natural or artificial impurities.
Such a system has numerous interesting features which will be
described in detail elsewhere \cite{EGST}; here we restrict ourselves
to a basic information.

For simplicity we set $\,d:=\pi\,$ and $\,\hbar^2/2m=1\,$, so that
in the absence of the impurities the motion is governed by the
Hamiltonian $\,H_0:= -\Delta\;$; the wavefunction is supposed to
satisfy the Dirichlet boundary conditions $\,\psi(x,0)=\psi(x,\pi)
=0\,$ for any $\,x\,$. Point interactions situated at $\,\vec a_j=
(a_j,b_j)\,, \; j=1,\dots,J\,$ can be introduced in the standard
way \cite[Sec.I.5]{AGHH}: they are determined by the boundary
conditions
$$
L_1(\psi,\vec a_j)+ 2\pi\alpha_j L_0(\psi,\vec a_j)\,=\,0\,,
\qquad j=1,\dots,J\,,
$$
relating the generalized boundary values
$$
L_0(\psi,\vec a)\,:=\, \lim_{|\vec x-\vec a|\to 0}\, {\psi(\vec
x)\over \ln |\vec x\!-\!\vec a|}\,, \quad\; L_1(\psi,\vec a)\,:=\,
\lim_{|\vec x-\vec a|\to 0} \Bigl\lbrack \psi(\vec x)\!-\! L_0(\psi,\vec
a)\, \ln |\vec x\!-\!\vec a| \Bigr\rbrack\,,
$$
where $\,\alpha_j\,$ are the (rescaled) coupling constants; the free
Hamiltonian corresponds to $\,\alpha_j=\infty\,,\; j=1,\dots,J\,$.

Solvability of the model stems from the fact that one can compute
the corresponding resolvent kernel by means of the Krein's formula; all
spectral information is contained then in the $\,J\times J\,$ matrix
$\,\Lambda(z)\,$ given by
  \begin{equation} \label{Lambda}
\Lambda(z)\,=\,\delta_{jm}\,\Big(\alpha_j-\xi(\vec a_j;z) \Big)-
(1\!-\!\delta_{jm})\,G_0(\vec a_j,\vec a_m;z)
   \end{equation}
where $\,G_0(\vec x_1,\vec x_2;z):= {{\rm i}\over\pi}\, \sum_{n=1}^{\infty}\,
\frac{{\rm e}^{{\rm i}k_n(z)|x_1-x_2|}} {k_n(z)}\, \sin(ny_1) \sin(ny_2)\,$ is the
free Green's function, $\,k_n(z):=\sqrt{z\!-\!n^2}\,$, and
  \begin{equation} \label{xi}
\xi(\vec a;z)\,:=\, \lim_{|\vec x-\vec a|\to 0}\,
\left(G_0(\vec a,\vec x;z)-
{1\over 2\pi}\,\ln|\vec x\!-\!\vec a| \right)\,=\,
{{\rm i}\over\pi}\,\sum_{n=1}^{\infty}\, \left(\,
{\sin^2(nb)\over k_n(z)}\,-\, {1\over 2{\rm i}n} \,\right)\,.
  \end{equation}
We suppose everywhere that the energy stays away of the thresholds,
$\,\sqrt z\ne 1,2,\dots\:$.
The knowledge of the resolvent allows us, in particular, to solve the
scattering problem. The reflection and transmission amplitudes are
expressed through the inverse of $\,\Lambda(z)\,$ as
   \begin{eqnarray} \label{rt}
r_{nm}(z) &\!=\!& {{\rm i}\over\pi}\, \sum_{j,k=1}^J \Lambda(z)^{-1}_{jk}
\,{\sin(mb_j) \sin(nb_k)\over k_m(z)}\, {\rm e}^{{\rm i}(k_ma_j+k_na_k)}\;,
\nonumber \\ \\
t_{nm}(z) &\!=\!& \delta_{nm}\,+\, {{\rm i}\over\pi}\, \sum_{j,k=1}^J
\Lambda(z)^{-1}_{jk}\, {\sin(mb_j) \sin(nb_k)\over k_m(z)}\,
{\rm e}^{-{\rm i}(k_ma_j-k_na_k)}\,. \nonumber
   \end{eqnarray}
The mirrored quantities are obtained by changing each perturbation
longitudinal coordinate $\,a_j\,$ to $\,-a_j\;$; together they satisfy
the unitarity condition
$$
\sum_{m=1}^{[\sqrt z]} k_m (t_{nm}\overline t_{sm}+r_{nm}\overline
r_{sm}) \,=\, \delta_{ns} k_n\,,\qquad
\sum_{m=1}^{[\sqrt z]} k_m \left(\tilde t_{nm}\overline r_{sm}
+\tilde r_{nm}\overline t_{sm}\right) \,=\,0 \,,
$$
where the summation runs over the open channels, $\,[\cdot]\,$
being the integer part.

\section{Comparison of the S--matrices}

From now on we shall consider a pair of point interactions, $\,J=2\,$,
with $\,a_1=0\,$ and $\,a_2=a\,$. For the sake of brevity we denote
$\,\gamma_j:=\alpha_j\!-\!\xi(\vec a_j;z)\,$ and $\,G:=
G_0(\vec a_1,\vec a_2;z)\,$, so the matrix $\,\Lambda(z)\,$ may
be written as {\scriptsize $\,\left(\begin{array}{cc} \gamma_1 & -G
\\ -G & \gamma_2 \end{array}\right)\,$}. For further purposes we
introduce also the truncated resolvent
   \begin{equation} \label{truncated resolvent}
G^t\,\equiv\, G_0^t(\vec a_1,\vec a_2;z):=\, {{\rm i}\over\pi}\,
\sum_{n=1}^M\,{\sin(nb_1)\,\sin(nb_2)\over k_n(z)}\, {\rm e}^{{\rm i}k_na}\;;
   \end{equation}
among the $\,M\,$ modes involved, $\,N:=[\sqrt z\,]\,$ are propagating
while the rest corresponds to evanescent channels. The one--perturbation
reflection and transmission are
   \begin{equation} \label{one point rt}
r^{(j)}_{nm}=\, {{\rm i}\over\pi}\,{\sin(nb_j)\sin(mb_j)\over k_m\gamma_j}\,
{\rm e}^{{\rm i}(k_n+k_m)a_j}\,,\quad
t^{(j)}_{nm}=\, \delta_{nm}+\,{{\rm i}\over\pi}{\sin(nb_j)\,\sin(mb_j)
\over k_m\gamma_j}\, {\rm e}^{{\rm i}(k_n-k_m)a_j}
   \end{equation}
for $\,j=1,2\,$. Suppose that we employ an $\,M\,$ component Ansatz
for wavefunctions in the intermediate region, $\,0<x<a\,$. Using the
definition (\ref{truncated resolvent}), we find
$$
\left(\,1-\tilde r^{(1)}r^{(2)}\right)_{nm}\,=\,
\delta_{nm} -\,{{\rm i}\over\pi}\:G^t\:{\sin(nb_1)\sin(mb_2)\over
k_m\gamma_1\gamma_2}\: {\rm e}^{{\rm i}k_m a}\,.
$$
This $\,M\times M\,$ matrix is explicitly invertible,
   \begin{equation} \label{inverse}
\left(\,1-\tilde r^{(1)}r^{(2)}\right)^{-1}_{nm}\,=\,
\delta_{nm} +\,{{\rm i}\over\pi}\:{G^t\over \gamma_1\gamma_2\!-\!(G^t)^2}\:
{\sin(nb_1)\sin(mb_2)\over k_m}\: {\rm e}^{{\rm i}k_m a}\,.
   \end{equation}
The relations (\ref{one point rt}) and (\ref{inverse}) allow us to
express the composed transmission amplitude as the lower left element
of the matrix (\ref{factorized S matrix}); after a straightforward
computation using the definition (\ref{truncated resolvent}), we
arrive at the formula
   \begin{eqnarray} \label{truncated transmission}
t^{(12)}_{nm} \,=\, \delta_{nm} &\!\!\!+\!\!& {{\rm i}\over\pi}\:{k_m^{-1}\over
\gamma_1\gamma_2\!-\!(G^t)^2}\, \bigg\{\, \gamma_2\sin(nb_1)\sin(mb_1)
+ G^t \sin(nb_2)\sin(mb_1)\,{\rm e}^{{\rm i}k_n a} \nonumber \\ \\
&& + G^t \sin(nb_1)\sin(mb_2)\,{\rm e}^{-{\rm i}k_m a}+
\gamma_1 \sin(nb_2)\sin(mb_2)\,{\rm e}^{{\rm i}(k_n-k_m)a}\, \bigg\}\,, \nonumber
   \end{eqnarray}
which differs from the exact expression (\ref{rt}) just by replacing
the {\em ``off--diagonal"} coefficients $\,G\,$ by the truncated one
$\,G^t\,$. A similar conclusion can be made for the other amplitudes in
(\ref{factorized S matrix}). Let us estimate the error due to neglecting
the remainder term
   \begin{equation} \label{remainder}
G^r\,:= {1\over\pi}\,
\sum_{n=M+1}^{\infty}\,{\sin(nb_1)\,\sin(nb_2)\over \sqrt{n^2\!-z}}\,
{\rm e}^{-a\sqrt{n^2\!-z}}\,.
   \end{equation}
Denote
$$
\tilde\gamma\,:=\,\max\{\,\gamma_1,\gamma_2\}\,, \qquad
\DD\,:=\, \sqrt{(M\!+\!1)^2\!-(N\!+\!1)^2}\,,
$$
where the maximum in the first expression is taken over the interval
of energies we are interested in. Since $\,N=[\sqrt z\,]\,$, we have
$$
(\ell\!+\!M\!+\!1)^2\!-z\,\ge\, (\ell\!+\!M\!+\!1)^2\!-(N\!+\!1)^2
\,\ge\, (\ell+\DD)^2\,,
$$
and the {\em relative} error of the ``diagonal" terms in
(\ref{truncated transmission}), \ie, the first and the last one, can
be estimated by
$$
|\Delta_{\rm d}|\,\aleq\, {2\over\pi}\,\left|\,{G\tilde\gamma\over
\gamma_1\gamma_2\!-\!(G^t)^2} \,\right|\:\sum_{n=M+1}^{\infty}\,
{{\rm e}^{-a\sqrt{n^2\!-z}}\over \sqrt{n^2\!-z}}\,\le\,
{2\over\pi\DD}\,\left|\,{G\tilde\gamma\over
\gamma_1\gamma_2\!-\!(G^t)^2} \,\right|\:
{{\rm e}^{-a\DD}\over 1\!-\!{\rm e}^{-a\DD}}\;;
$$
for the off-diagonal ones we have
$$
|\Delta_{\rm off}|\,\aleq\, {1\over\pi\DD}\,
\left|\,{\gamma_1\gamma_2\!+\!G^2\over \gamma_1\gamma_2\!-\!(G^t)^2}
\,\right|\:{{\rm e}^{-a\DD}\over 1\!-\!{\rm e}^{-a\DD}}\,.
$$

This error bounds show where deviations from the exact expression
are most likely. An obvious requirement is that the evanescent
states must be given opportunity to decay; as long as the moduli
are of order of one, it is the exponential term which governs the
estimates. To get a $\,K\,$ digit precision, one roughly needs
$\,(M\!+\!1)^2\!-(N\!+\!1)^2\ageq \left(K\over 2a\right)^2\,$.
In general, the problem becomes therefore non--trivial when we glue
scatterers without intermediate regions; then one has to check that
the evanescent states do indeed decay within each $\,S_j\,$ when we
move from its centre to the boundaries.

The second source of error are the denominators. The two--impurity
system has no embedded eigenvalues (apart from the trivial ones due
to symmetry), so the exact $\,\gamma_1\gamma_2\!-\!G^2\,$ is never
zero in the cases of interest \cite{EGST}. On the other hand, it has
resonances unless the impurity sits at a node of a transverse
eigenfunction; they are narrow in the case of a weak coupling, \ie,
for $\,\alpha_j\,$ large positive. The truncated expression may blow
up around the resonance energies, and even if it is not the case, it
may produce shifted resonances. This is important, because resonance
structures such as conductivity modulations around thresholds are a
primary object of interest in quantum--wire serial scatterers.

To illustrate the differences due to the cut--off factorization, let
us compute the conductivity of the quantum wire with a pair of
impurities,
   \begin{equation} \label{conductivity}
g(z)\,=\, {2{\rm e}^2\over h}\, \sum_{n,m=1}^{[\sqrt z\,]}\,
{k_m\over k_n}\, |t_{nm}(z;M)|^2\,,
   \end{equation}
where $\,t_{nm}(z;M)\,$ is given by (\ref{truncated transmission}) with
the resolvent (\ref{truncated resolvent}) which includes $\,M\!-\!N\,$
evanescent states; the limit $\,M\to\infty\,$ gives the exact answer.
Let us remark that the one--point problem contains also a sum over all
modes in the function (\ref{xi}) but this can be controlled \cite{EGST}.
We see that even if the approximations with low number of evanescent
modes \ie $\,M \aleq 5$ differ considerably only in the vicinity of the
thresholds, there are regions like those close to a resonance (\cf the
inset), where the convergence slows rapidly down. For instance, in order
to get the correct position of the peak, higher evanescent modes, up to
$M=15$ in this particular case, must be taken into account.

\section*{Acknowledgement}
The work has been partially supported by the Grants AS No.148409
and GA CR No.202--93-1314.

\newpage

\noindent{\bf Figure caption:}

\bigskip
\bigskip
\bigskip

\noindent
Figure 1: A pair of scatteres.

\bigskip
\bigskip
\bigskip

\noindent
Figure 2: A pair of point scatteres in the strip.

\bigskip
\bigskip
\bigskip

\noindent
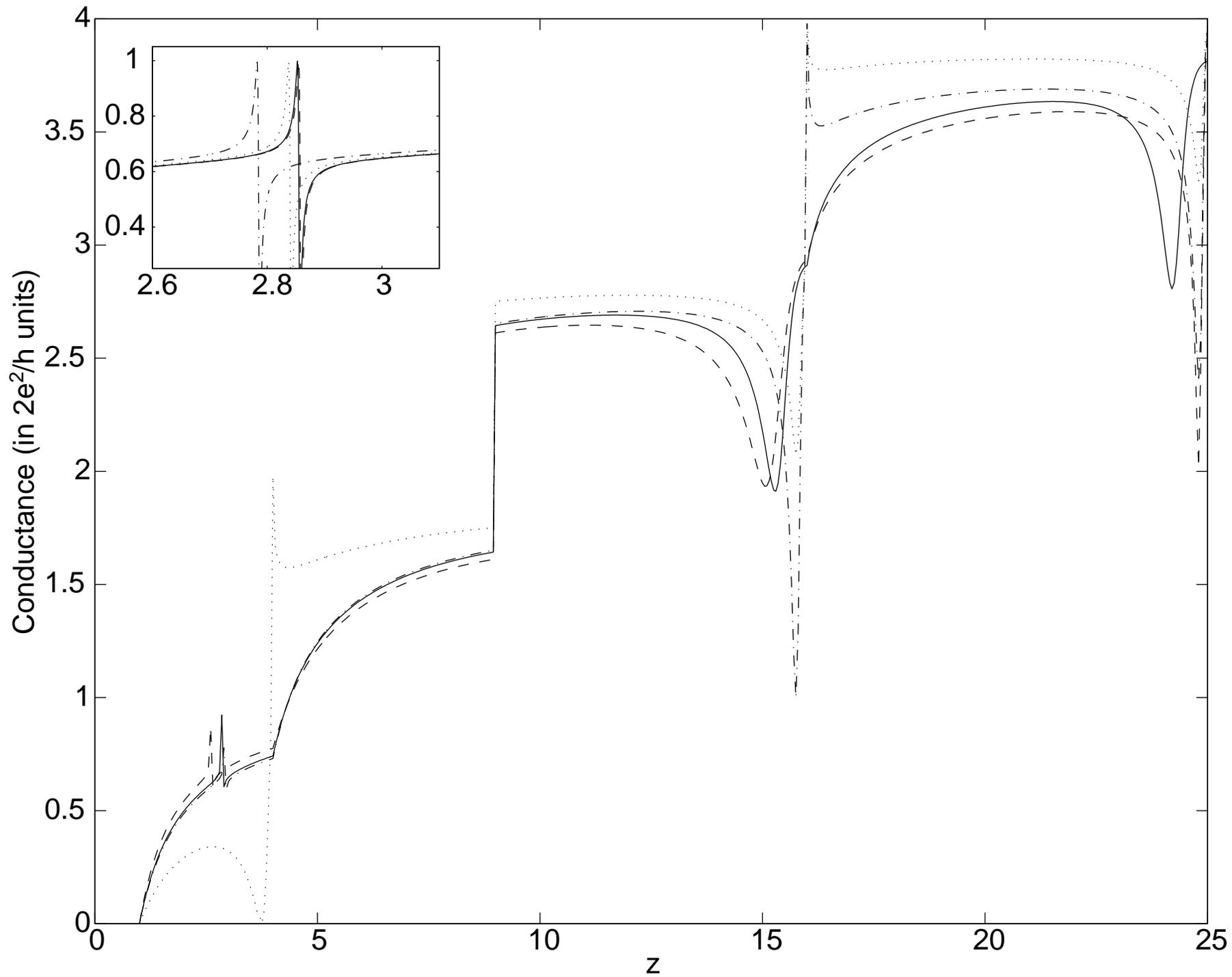
Figure 3: Conductivity plots for factorized {\em vs.} exact
scattering matrices. The positions of scatterers are $\,\vec a_1=
(0,\pi/3),\, \vec a_2=(0.05,2\pi/3)$ and the coupling constants
are chosen $\alpha_1=\alpha_2=0.2$.
The full line represents the exact result,
dotted, dash-dotted and dashed lines the approximations with
$M=1$, $M=2$ and $M=4$, respectively. In the inset, however, they
represent the approximations with $M=5$, $M=10$ and $M=15$,
respectively.

   \end{document}